\documentclass[trackchanges]{aastex701}

\usepackage{amsmath}
\usepackage{threeparttable}

\begin{document}

\title{SED and Galactic kinematic diagnostics for dormant BH/NS binary candidates} 

\correspondingauthor{Wei-Min Gu}
\email{guwm@xmu.edu.cn}

\author[0009-0009-3791-0642]{Qian-Yu An}\email{anqianyu@stu.xmu.edu.cn}
\affiliation{Department of Astronomy, Xiamen University, Xiamen, Fujian 361005, People's Republic of China} 

\author[0000-0003-3137-1851]{Wei-Min Gu}\email{guwm@xmu.edu.cn}
\affiliation{Department of Astronomy, Xiamen University, Xiamen, Fujian 361005, People's Republic of China}

\begin{abstract}
The third data release of the \emph{Gaia} mission (\emph{Gaia} DR3) has enabled large-scale searches for dormant black hole and neutron star binaries with stellar companions at AU-scale separations. A recent study has proposed thousands of dormant black hole and neutron star binary candidates using summary statistics from \emph{Gaia} DR3 by simulating and fitting \emph{Gaia} observables. In this work, we perform broadband spectral energy distribution (SED) fitting from the optical to the infrared for 1,328 candidates, incorporating GALEX ultraviolet photometry to assess the presence of hidden hot companions. We quantify ultraviolet excess by comparing observed near-ultraviolet fluxes with single-star SED predictions and further test whether excesses can be explained by non-degenerate stellar companions for sources exhibiting moderate excess. We additionally examine the Galactic kinematics of the sample to identify systems potentially affected by natal kicks during compact-object formation. By combining the ultraviolet and kinematic diagnostics, we identify 182 sources as the highest-priority candidates for follow-up observations, in which 19 are black hole candidates with \texttt{fit\_companion\_mass} $\geq$ 3 $M_\odot$.

\end{abstract}

\keywords{\uat{Binary stars}{154} --- \uat{Black holes}{162} --- \uat{Neutron stars}{1108} --- \uat{Stellar kinematics}{1608} --- \uat{Spectral energy distribution}{2129}}

\section{Introduction} 
Theoretical models suggest the Milky Way harbors billions of black holes (BHs) and neutron stars (NSs) \citep[e.g.,][]{1996ApJ...457..834T, 2010A&A...510A..23S}, driving decades-long searches for these objects. Historically, accreting BH X-ray binaries have been primarily discovered since the 1960s through their characteristic X-ray bursts, leading to the identification of $\sim$30 accreting BH X-ray binaries \citep[e.g.,][]{2016A&A...587A..61C}. NS searches have employed both X-ray bursts and radio pulse detections, yielding hundreds of detections via X-ray bursts \citep[e.g.,][]{2024A&A...684A.124F} and thousands via radio pulses \citep{2005AJ....129.1993M}. 
\par However, these classical methods sampled only a small fraction of the total BH and NS population. X-ray burst detections prefer accreting BHs/NSs, while radio pulse detections can only detect rapidly rotating NSs which emit electromagnetic pulse signals beaming towards us. Recent technological advances and data accumulation have spurred the development of novel search techniques. Microlensing remains the only currently feasible method for detecting isolated stellar-mass BHs, yielding several candidates \citep[e.g.,][]{2016MNRAS.458.3012W} and one confirmed detection in the event OGLE-2011-BLG-0462 \citep{2022ApJ...933L..23L,2023ApJ...955..116L,2025ApJ...983..104S}. Future Galactic Bulge Time Domain Survey \citep{2019ApJS..241....3P} using the Nancy Grace Roman Space Telescope, formerly called Wide Field Infrared Survey Telescope \citep{2019arXiv190205569A}, is expected to discover many more such isolated BHs/NSs. With the detection of GW150914 \citep{2016PhRvL.116f1102A}, gravitational wave observations are dedicatedly revealing merging BH-BH and BH-NS binaries, and more than 300 events have been detected \citep{2025arXiv250818079T}. Large-scale spectroscopy surveys have amassed vast stellar spectral datasets, enabling the identification of non-accreting BHs and NSs in medium-to-short orbital period binaries through radial velocity (RV) monitoring \citep[e.g.,][]{2018MNRAS.475L..15G, 2019A&A...632A...3G, 2022NatAs...6.1085S, 2022A&A...664A.159M, 2022NatAs...6.1203Y}. Furthermore, photometry has also been adopted to search for dark massive companions (BHs/NSs) in ellipsoidal variables \citep[e.g.,][]{2021MNRAS.501.2822G, 2023A&A...674A..19G, 2024OJAp....7E..24R, 2025A&A...695A.210G}.
\par While discoveries from these novel methods have expanded the known populations of BHs and NSs, they have yet to significantly increase their total known numbers. Importantly, the third data release of the \emph{Gaia} mission \citep[\emph{Gaia} DR3;][]{2023A&A...674A...1G} offers a promising new avenue to overcome this limitation. Its initial data release included orbital solutions for 1.7$\times$10$^{5}$ binaries with full astrometric parameters. Utilizing these full astrometric solutions along with spectroscopic observations, BHs and NSs in AU-scale-orbit binaries are being revealed \citep[e.g.,][]{2023MNRAS.518.1057E, 2023MNRAS.521.4323E}. In addition, several BH binaries associated with \emph{Gaia} astrometric measurements have also been reported \citep{2024A&A...686L...2G, 2024NatAs...8.1583W}. However, only a small fraction of AU-scale-orbit binaries have such published data due to the stringent quality cuts, inevitably missing detections for many BHs and NSs in AU-scale-orbit systems. To address this, \citet{2025arXiv251005982M} enlarged their search to the extended \emph{Gaia} DR3 dataset. They employed a forward-modeling framework, simulating \emph{Gaia} observables for 21,028 red-giant branch (RGB) stars to target massive dark companions, yielding 556 RGB + BH candidates. 
\par Assessing the risk of hidden hot companions is necessary for these candidates, given previous false alarms like J0521 and V723 Mon. Both systems were initially reported as RGB + BH binaries \citep{2019Sci...366..637T, 2021MNRAS.504.2577J}. However, later analyses revealed that they host stellar companions rather than BHs: ultraviolet (UV) features indicated a hot companion in J0521 \citep{2024ApJ...976..131B}, while for V723 Mon, spectral disentangling already demonstrated the presence of a stellar companion \citep{2022MNRAS.512.5620E}, with subsequent UV observations providing additional confirmation \citep{2025arXiv250910608K}. While UV excess provides a powerful diagnostic for identifying hot luminous stellar contaminants, kinematic signatures offer an independent probe of compact-object formation. Supernova explosions associated with the birth of BHs or NSs can impart substantial natal kicks to binary systems through asymmetric mass loss and explosion kinematics, leading to significantly heated binary orbits and high peculiar velocities ($V_{\rm pec}$) \citep[e.g.,][]{1961BAN....15..265B, 1994Natur.369..127L, 2005MNRAS.360..974H, 2013MNRAS.434.1355J}. Kinematic information therefore provides a complementary avenue to distinguish genuine compact-object binaries from ordinary stellar systems.
\par In this work, we perform spectral energy distribution (SED) analysis for 1,328 sources from \citet{2025arXiv251005982M}, along with Galactic kinematic analysis, as a reference for evaluating their reliability and priority selection of follow-up observation sources. The structure of this paper is organized as follows. In Section \ref{sec2}, we introduce the source selection for analysis in this work and corresponding data collection. In Section \ref{sec3}, we introduce our SED fitting method and use UV diagnostics to select sources. In Section \ref{sec4}, we analyze the Galactic kinematics of our sample. We summarize our results and make a discussion in Section \ref{sec5}.

\section{source selection and data collection}\label{sec2}
Our initial sample comprises 3,773 sources with \texttt{fit\_companion\_mass} $\geq$ 1.4 $M_\odot$ and \texttt{flag\_quality == True} from \citet{2025arXiv251005982M}. Due to the necessity of UV diagnostics to evaluate the risk of hidden hot companions, we refine sources by requiring available UV photometry. The Galaxy Evolution Explorer (GALEX) \citep{2005ApJ...619L...1M, 2005ApJ...619L...7M} is uniquely useful for this purpose, being the only modern mission that performed a large-area, UV photometric imaging survey of the sky with high spatial resolution. Equipped with a 50-cm Ritchey–Chr{\'e}tien telescope, GALEX can simultaneously observe in two broadband channels—the far-ultraviolet (FUV; 1,350–1,750 $\mathrm{\AA}$) and near-ultraviolet (NUV; 1,750–2,750 $\mathrm{\AA}$). GALEX General Releases 6 and 7 (GR6/GR7), which constitute the final public data products of the mission and are based on the GALEX Merged Catalog of Sources \citep[MACT;][]{https://doi.org/10.17909/t9h59d}, contain 82,992,086 sources. We cross-match our sample with GALEX GR6/GR7 to acquire UV photometry, reducing the sample to 903 sources. We further retrieve the MCAT (\texttt{AIS\_$\ast$\_mcat.fits}) by python codes \texttt{from astroquery.mast import Catalogs} and \texttt{Catalogs.query\_region()} \citep{2019AJ....157...98G}, which provides UV photometry for another 645 sources.
\par To construct the broadband SED, we supplement photometry from existing \emph{Gaia} G, BP and RP bands by retrieving the archive of The AAVSO Photometric All-Sky Survey (APASS) \citep[$g$, $r$, $i$, $B$ and $V$ bands;][]{2015AAS...22533616H}, Two Micron All-Sky Survey (2MASS) \citep[$J$, $H$ and $K$ bands;][]{2006AJ....131.1163S} and Wide-field Infrared Survey Explorer \citep[$W1$ and $W2$ bands;][]{2010AJ....140.1868W}, to collect optical and infrared photometry. Photometric measurements with undefined or zero uncertainties are regarded as invalid measurements. Furthermore, for photometric measurements from APASS, we also regard photometric measurements with \texttt{Uncertainty flag = 1} as invalid measurements. Sources retrieved in APASS may only have valid photometry for partial bands, and we only retain sources with at least three bands of valid photometry to ensure enough photometric data points. Finally, we only consider sources with \texttt{parallax\_over\_err $>$ 5}, leaving 1,349 sources.
\par Given the inherent degeneracies in SED fitting, incorporating priors on stellar parameters is essential. We adopt stellar parameters published by \citet{2025ApJ...984...58H}, who have developed a new pipeline, \emph{Gaia} Net, for reprocessing \emph{Gaia} XP spectra, to predict stellar parameters for 220 million stars released in \emph{Gaia} DR3 and published their results on \texttt{Vizier}\footnote{\url{https://vizier.cds.unistra.fr/viz-bin/VizieR?-source=J/ApJ/984/58}} \citep{2000A&AS..143...23O, 2025yCat..19840058H}, due to the complete sample coverage and accessibility. 21 sources without uncertainties for all three stellar parameters have been ruled out in this process, leaving 1,328 sources for the following analysis. 

\section{SED fitting}\label{sec3}
We set an additional systematic photometric uncertainty of 0.03 mag as an uncertainty floor for each band, then we fit the SEDs with a single-star model. Since our intention is to conduct UV excess diagnostics, we do not include the GALEX photometry in SED fitting. The parameters for SED fitting include effective temperature $T_\mathrm{eff}$, surface gravity $\log g$, metallicity [Fe/H], stellar radius $R$, distance $D$, and $V$-band extinction $A_{\rm V}$. We perform the SED fitting using an affine-invariant Markov Chain Monte Carlo sampler implemented in \texttt{emcee} \citep{2013PASP..125..306F} to explore the posterior distribution of the model parameters. We run 24 walkers for 5,000 steps, regarding the initial 2,500 steps of each walker as the burn-in and discarding them. The posterior probability of the fitting parameters $P_\mathrm{rob}$ is
denoted as:
\begin{equation}
 \ln P_{\rm rob} = \ln \mathcal{L} + \ln \mathcal{P},   
\end{equation}
where $\mathcal{L}$ is the likelihood and $\mathcal{P}$ is prior. We adopt separable priors for all model parameters. Gaussian priors are applied to ($T_{\rm eff}$, $\log g$, [Fe/H]) and $D$, based on \emph{Gaia} Net estimates and the \emph{Gaia} parallax, respectively, with widths set by the quoted uncertainties of these quantities. $A_\mathrm{V}$ and $R$ are assigned non-informative flat priors over their physically allowed ranges. The likelihood $\mathcal{L}$ is denoted as:
\begin{equation}
 \ln L = -\frac{1}{2} \sum_\mathrm{band}^{} \left(\frac{F_{\rm obs,band}-F_{\rm mod,band}(\emph{T}_\mathrm{eff}, \log g, {\rm [Fe/H]}, \emph{R}, \emph{D}, \emph{A}_{\rm V})}{\sigma_{\rm obs}}\right)^{2}\,
\end{equation}
where $F_{\rm obs,band}$ is the observed flux, $F_{\rm mod,band}$ is the model flux, and $\sigma_{\rm obs}$ is the error of the observed flux. For each set of fitting parameters, the model flux of each band is calculated by convolving the synthetic spectrum with the corresponding transmission curve provided by the virtual observatory SED analyzer\footnote{\url{https://svo2.cab.inta-csic.es/theory/fps/}} (VOSA) \citep{2008A&A...492..277B}. The synthetic spectrum is generated by \texttt{pystellibs}\footnote{\url{https://github.com/mfouesneau/pystellibs}} for a given set of ($T_\mathrm{eff}$, $\log g$, [Fe/H], $R$), using \texttt{BTSettl} library \citep{2001ApJ...556..357A, 2012RSPTA.370.2765A, 2016sf2a.conf..223A}, then reddened according to the Fitzpatrick extinction law \citep{1999PASP..111...63F}. The total-to-selective extinction ratio $R_\mathrm{V}$ is set to be 3.1 \citep{1989ApJ...345..245C} and the reddening coefficients at different wavelengths for a given $A_\mathrm{V}$ are calculated by \texttt{extinction} \citep{2016zndo....804967B}. Finally, the synthetic spectrum is diluted by 4$\pi D^2$. The zero-point used for converting magnitude to flux of each band is also provided by VOSA.   
\par For each source, we adopt the median of the posterior distribution for each parameter as the best-fitting value and generate the corresponding synthetic spectrum. We then compute the \( \chi^2 \) per data point (\( \chi^2 \)/$N$) of the fit using all non-UV photometric bands. To ensure that the observed broadband photometry is adequately described by a single-star SED model, we retain only sources whose \( \chi^2 \)/$N$ $<$ 2. As a result, 137 sources are excluded, yielding a cleaner sample containing 1,191 sources for subsequent investigations.
\par
\begin{figure*}[ht!]
\plotone{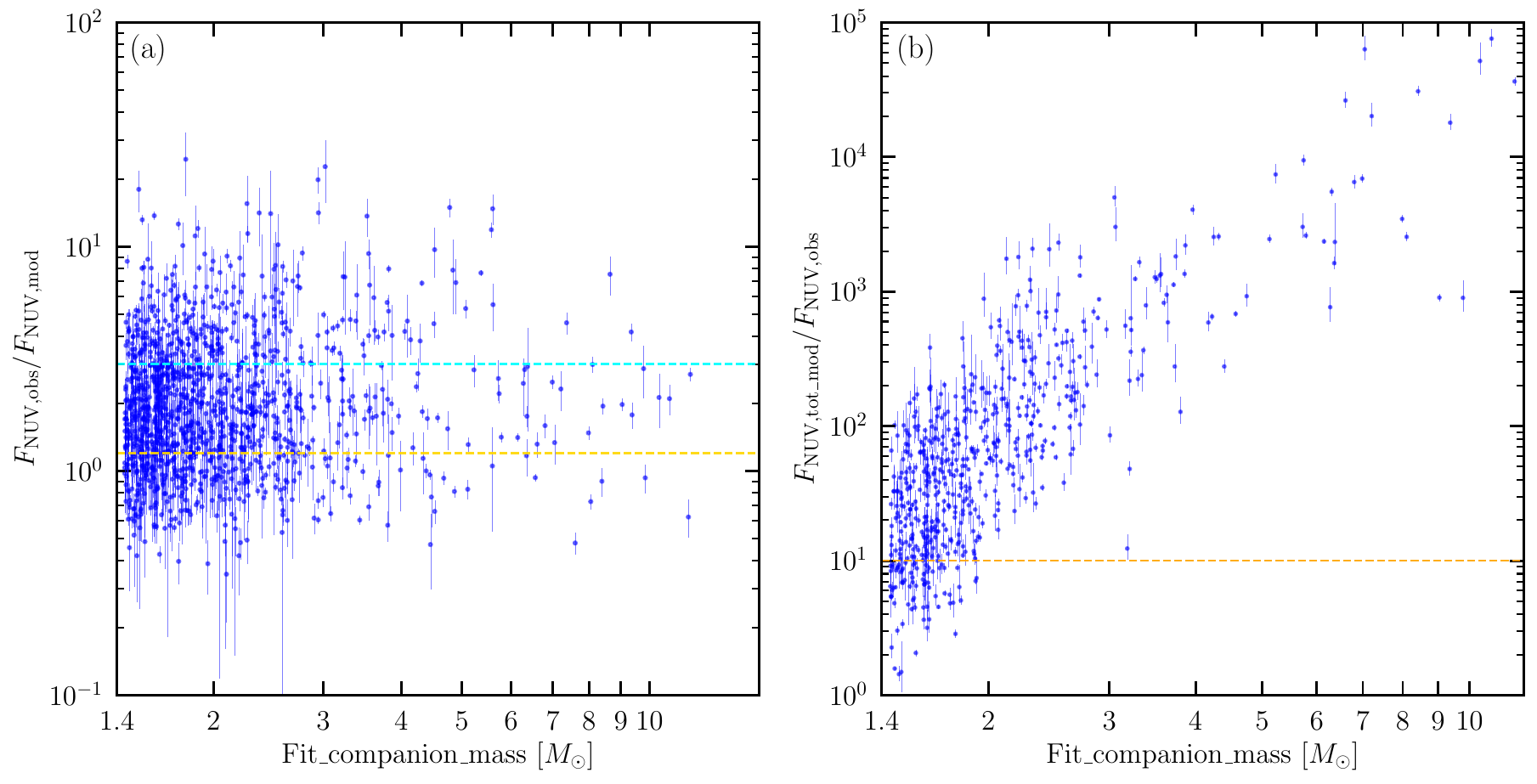}
\caption{{\bf (a), The ratio of observed NUV flux to model-predicted NUV flux as a function of the \texttt{fit\_companion\_mass}}. The horizontal orange dashed line marks $F_\mathrm{NUV,obs}$/$F_\mathrm{NUV,mod}=1.2$, below which sources are considered to show no significant NUV excess and are directly retained. Sources with moderate excess, $1.2<F_\mathrm{NUV,obs}$/$F_\mathrm{NUV,mod}\leq3$, indicated by the area between the cyan dashed line and orange dashed line, are subjected to further analysis allowing for a luminous stellar companion. {\bf (b), The ratio of total model-predicted NUV flux (assuming MS companion) to observed NUV flux as a function of the \texttt{fit\_companion\_mass}. The horizontal orange dashed line marks $F_\mathrm{NUV,tot\_mod}$/$F_\mathrm{NUV,obs}=10$, above which the predicted flux from a non-degenerate luminous companion severely exceeds the observed value, making such companions less possible.}
\label{fig:SED_diagnostics}}
\end{figure*}

\begin{figure*}[ht!]
\plotone{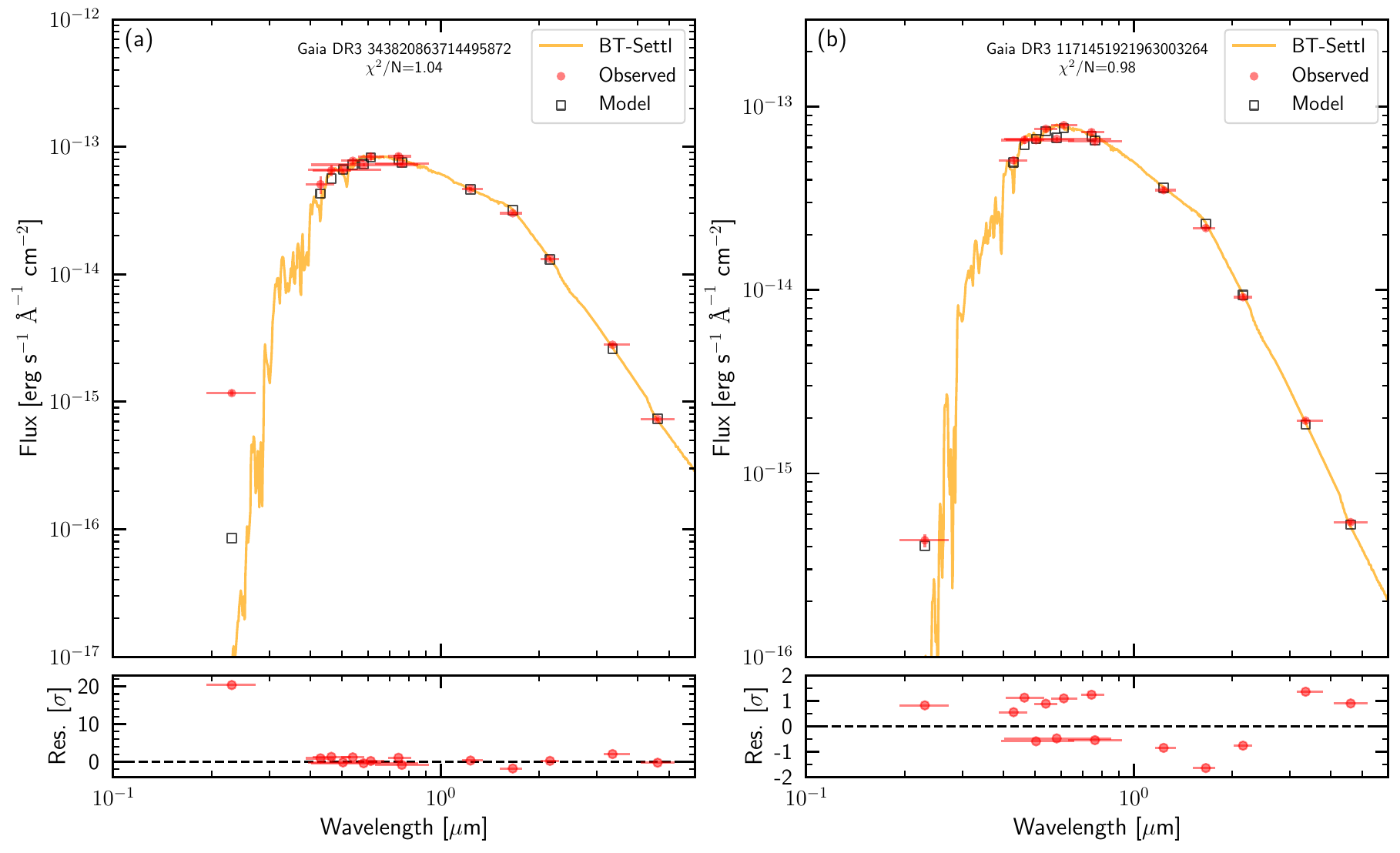}
\caption{{\bf (a), SED example for a candidate with a UV excess. (b), SED example for a candidate without a UV excess.} The optical-to-infrared SEDs of both sources can be well fitted by a single-star SED model. For each panel, red points show observed fluxes, open black squares show model-predicted fluxes, and the orange curve shows the best-fit spectrum.
\label{fig:SED_example}}
\end{figure*}

 For the retained sources, we compute the model-predicted NUV fluxes and compare them with the observed values (see Figure \ref{fig:SED_diagnostics} (a)) and show SED examples for a candidate with a UV excess and one without a UV excess, respectively, in Figure \ref{fig:SED_example}. We define the NUV flux ratio as $R_{\rm NUV}=F_\mathrm{NUV,obs}/F_\mathrm{NUV,mod}$. 298 sources with $R_{\rm NUV}\leq1.2$ are considered to show no significant NUV excess and are directly retained. We note that some sources have $R_{\rm NUV}<1$, which is not physically expected and likely reflects measurement uncertainties or limitations of the model; no additional filtering is applied to these objects. For 585 sources with moderate excess, $1.2<R_{\rm NUV}~\leq~3$, we perform further analysis to test whether the observed NUV emission can be explained by a non-degenerate companion or not.
\par To test this hypothesis, we model the companion in the moderate-excess systems as a main-sequence (MS) star. Stellar parameters for the companion are inferred from \texttt{Isochrones}\footnote{\url{https://isochrones.readthedocs.io/en/latest/index.html}} \citep{2015ascl.soft03010M}, adopting the same [Fe/H] as the RGB in each system. The predicted parameters are used to generate synthetic spectra, which are then reddened using the SED-derived extinction and scaled by the distance to predict the observed fluxes and compute the corresponding NUV fluxes. We then compare the total model-predicted NUV fluxes of the RGB and the companion with the observed values (see Figure \ref{fig:SED_diagnostics} (b)). We find a strong positive correlation between the predicted-to-observed NUV flux ratio and the \texttt{fit\_companion\_mass}, as expected if the companion were a normal luminous star. For a substantial fraction of sources, the total model flux exceeds the observed NUV flux by more than an order. Such large discrepancies are difficult to reconcile with measurement uncertainties or model limitations, indicating that a non-degenerate luminous companion is less possible. We therefore interpret 512 systems with $F_\mathrm{NUV,tot\_mod}/F_\mathrm{NUV,obs}\geq10$ as more consistent with hosting dark companions and retain them as candidate BH/NS binaries. After applying both the SED quality selection and the UV consistency tests, 810 sources satisfy the SED diagnostics.

\section{Galactic kinematic diagnostics}\label{sec4}
We characterize the kinematic properties of 1,191 sources, whose optical-to-infrared SEDs can be well fitted by single-star model, by deriving their $V_{\rm pec}$ and maximum vertical heights above/below the Galactic plane ($|Z|_\mathrm{max}$) in the Galactic potential. We evaluate the $V_{\rm pec}$ at the Galactic-plane crossing phase ($Z=0$). This choice provides a uniform kinematic reference across the sample, where the local circular velocity is well defined in the adopted Galactic potential model, and simultaneously reduces sensitivity to orbital-phase–dependent exchange between kinetic and gravitational potential energies. Systemic velocities needed for $V_{\rm pec}$ calculation are taken from \texttt{radial\_velocity} published by \emph{Gaia} DR3, which are derived from multi-epoch RV observations and provided as a robust average value for each source. Therefore, the use of \emph{Gaia} DR3 \texttt{radial\_velocity} as systemic velocities is reasonable. We integrate the orbits backward for 1 Gyr in the Galactic potential under the Milky Way potential model \texttt{McMillan17} \citep{2017MNRAS.465...76M}, the most used Milky Way potential model, implemented by python package \texttt{galpy} \citep{2015ApJS..216...29B}, to determine the Galactic plane crossing phase and $|Z|_\mathrm{max}$. Then we calculate the 3D velocities in Galactocentric Cartesian frame ($V_{\rm x}$, $V_{\rm y}$, and $V_{\rm z}$) at the Galactic plane crossing phase. In the calculation process, we place the Sun at (\emph{X}, \emph{Z}) = (-8.178, 0.025) kpc \citep{2016ARA&A..54..529B, 2019A&A...625L..10G}. At such place, the circular velocity is set to be 233.2 km s$^{-1}$ according to \texttt{McMillan17}. $V_{\rm pec}$ of the Sun relative to the Local Standard of Rest is set to be ($U_\odot$, $V_\odot$, $W_\odot$) = (7.01, 10.13, 4.95) km s$^{-1}$ \citep{2015MNRAS.449..162H}. We convert $V_{\rm x}$, $V_{\rm y}$, and $V_{\rm z}$ into 3D Galactic space velocity ($U,~V,~W$) by the following matrix transformation:

$$
\begin{bmatrix}
    -\cos\alpha & -\sin\alpha & 0 \\
    -\sin\alpha & \cos\alpha & 0 \\ 0 & 0 & 1    
\end{bmatrix}
\begin{bmatrix}
    V_{\rm x} \\
    V_{\rm y} \\
    V_{\rm z}
\end{bmatrix}
=
\begin{bmatrix}
    U \\
    V \\
    W
\end{bmatrix}
,$$
where $\alpha$ is the angle between the Galactic Center-to-source vector and the positive X-axis in Galactocentric Cartesian frame. From the resulting Galactic velocity components, we define the $V_{\rm pec}$ as $V_\mathrm{pec}=\sqrt{U^{2} + (V-V_{\rm c})^{2} + W^{2}}$, where $V_{\rm c}$ is the local circular velocity at the disk-crossing radius.
\par \citet{2026MNRAS.tmp...46Z} investigated $V_{\rm pec}$ of compact-object binaries and major contaminating populations. By comparing their empirical cumulative distributions, they showed that the $V_{\rm pec}$ of more than 90\% of contaminating sources lie below 100 km s$^{-1}$ (see their Figure 3). We therefore adopt $V_{\rm pec}\geq100$ km s$^{-1}$ to isolate systems probably accelerated by supernova natal kicks, yielding 227 sources. Among them, 182 sources also satisfy the SED-based selection criteria and 19 are BH candidates with \texttt{fit\_companion\_mass} $\geq3$ $M_\odot$. On the other hand, theoretical studies suggest that BHs may experience substantially reduced natal kicks due to strong fallback \citep[e.g.,][]{2024MNRAS.527.8586T} or even form via direct collapse with negligible mass ejection \citep[e.g.,][]{1999ApJ...522..413F}, naturally leading to lower $V_{\rm pec}$. Observations also support that some BHs must form with very weak kicks \citep[e.g.,][]{2025PASP..137c4203N}. Therefore, 77 sources that satisfy the SED criterion alone and have \texttt{fit\_companion\_mass} $\geq3$ $M_\odot$ remain plausible BH binary candidates.
\par
\begin{figure*}[ht!]
\plotone{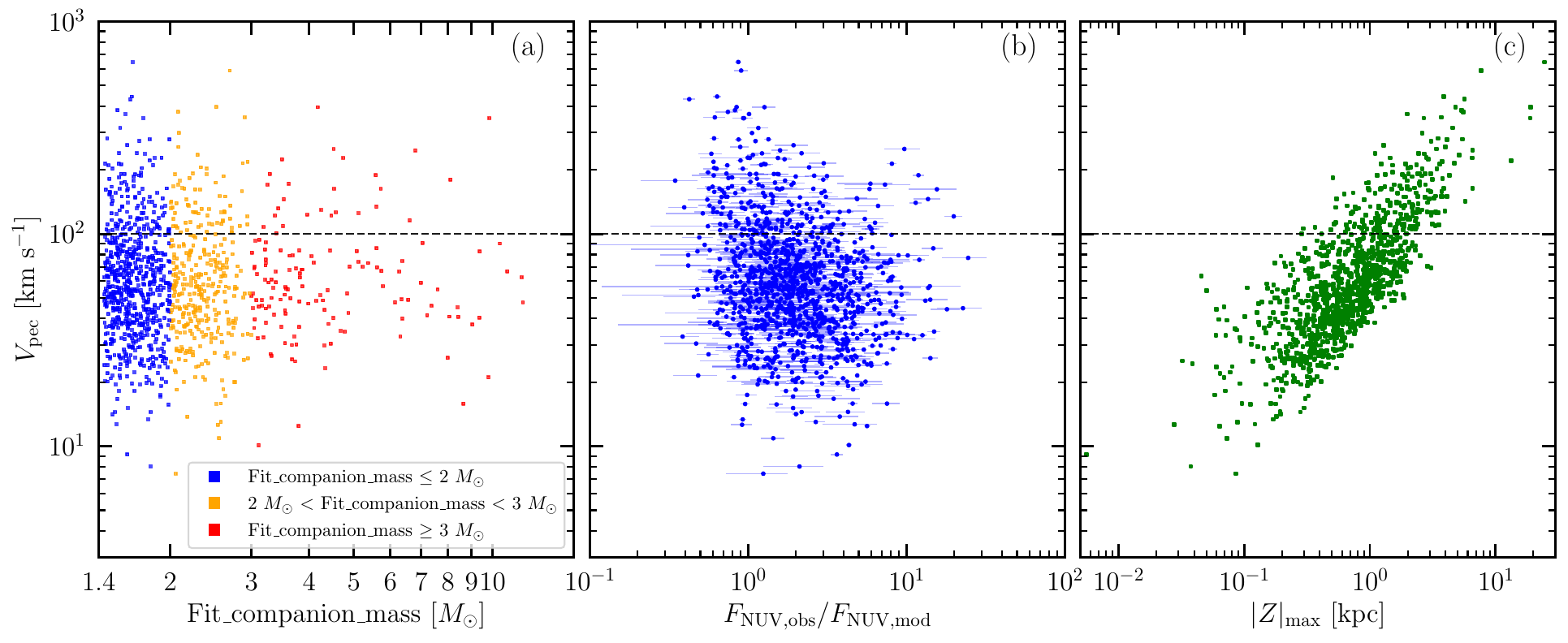}
\caption{{\bf (a),} $V_{\rm pec}$ as a function of the \texttt{fit\_companion\_mass}. NS-like zone, transition zone, and BH-like zone are respectively indicate by blue squares, orange squares and red squares. {\bf (b),} $V_{\rm pec}$ as a function of $R_{\rm NUV}$. {\bf (c),} $V_{\rm pec}$ as a function of $|Z|_{\rm max}$. In all panels, the horizontal black dashed line marks $V_{\rm pec}$ = 100 km s$^{-1}$, above which sources are considered to be probably accelerated by supernova natal kicks. For NS-like zone, transition zone, and BH-like zone, there are 123, 68, and 25 systems that satisfy this criterion, respectively.
\label{fig:peculiar_mass_nuv_ratio}}
\end{figure*}

 In Figure \ref{fig:peculiar_mass_nuv_ratio}, we respectively compare $V_{\rm pec}$ with \texttt{fit\_companion\_mass}, $R_{\rm NUV}$, and $|Z|_{\rm max}$, for sources with \( \chi^2 \)/$N$ $<$ 2. According to the \texttt{fit\_companion\_mass}, we divide the systems into NS-like zone, transition zone, and BH-like zone, respectively corresponding to \texttt{fit\_companion\_mass} of $\leq2$ $M_\odot$, $<2$ $M_\odot$ and $<3$ $M_\odot$, and $\geq3$ $M_\odot$. These three zones contain 677, 322, and 134 sources, respectively. The NS-like zone extends to higher $V_\mathrm{pec}$ tails than the BH-like zone; however, given the much larger sample size in the low-mass bin and possible selection effects, our data do not permit a quantitative comparison of natal kick distributions between the two populations. We further note a mild tendency for systems with higher $V_{\rm pec}$ to exhibit lower NUV flux ratios, with the median $R_{\rm NUV}$ shifting from $\sim$2 at $V_{\rm pec}< 100$ km s$^{-1}$ to $\sim$1.25 at higher $V_{\rm pec}$. This behavior is consistent with kinematically heated systems preferentially hosting compact companions, while more significant NUV excess is more common among kinematically cold populations. Interestingly, the median $R_{\rm NUV}$ shifting in the subsample with $V_{\rm pec}\geq100$ km s$^{-1}$, $\sim$1.25, matches the threshold adopted in Section \ref{sec3} for identifying sources showing no significant NUV excess, providing an independent consistency check between the UV and kinematic diagnostics. Furthermore, a clear positive correlation is observed between $V_{\rm pec}$ and $|Z|_{\rm max}$, indicating progressively heated orbits from disk-like to halo-like kinematics.

\section{Summary and discussion}\label{sec5}
\citet{2025arXiv251005982M} employed a forward-modeling framework, simulating \emph{Gaia} observables for 21,028 RGB stars to target compact object candidates. According to their results, we perform SED and Galactic kinematic diagnostics to provide additional vetting for 1,328 dormant BH/NS binary candidates and to prioritize targets for follow-up observations. A total of 810 sources satisfy the SED-based selection criteria. Their optical and infrared photometry can be well fitted by a single-star model and exhibit no UV excess or moderate UV excess that is completely insufficient to accommodate the existence of non-degenerate companions. Independent of the SED-based diagnostics, 227 sources meet the kinematic criterion with $V_\mathrm{pec}\geq100$ km s$^{-1}$. We identify the intersection of these two samples as the highest-priority candidates for follow-up observations. The highest-priority candidates contain 182 sources, of which 19 are BH candidates with \texttt{fit\_companion\_mass} $\geq$ 3 $M_\odot$. Besides, we designate 77 sources that satisfy the SED criterion alone and have \texttt{fit\_companion\_mass} $\geq$ 3 $M_\odot$ as a secondary-priority sample which only contains BH candidates. Both the highest-priority and secondary-priority candidates are listed in Table~\ref{tab:candidates}. 
\par Hierarchical systems may contaminate our sample, as their AU-scale orbits can accommodate inner binaries. In addition, for systems with moderate NUV excess but unlikely to host MS companions, contamination may arise from evolved non-degenerate companions in evolutionary phases where the NUV flux is significantly reduced. Based on \texttt{Isochrones}, we estimate that this phase occupies a non-negligible but limited fraction (typically $\sim$14\% to $\sim$32\%). Notably, such scenarios are only relevant to the subset of sources with moderate NUV excess, which accounts for $\sim$63\% of the systems that satisfy the SED diagnostics, suggesting that the overall impact of such contamination is further limited. For highest-priority candidates, the high $V_{\rm pec}$ indicates a comparatively higher likelihood of having experienced supernova explosions. Consequently, the probability of such systems being contaminated by evolved non-degenerate companions becomes lower. Furthermore, even if such systems host inner binaries, they are expected to contain at least one compact object (BH or NS) rather than being composed purely of non-degenerate stars. In contrast, the secondary-priority sample, with lower $V_{\rm pec}$, is more likely to be contaminated by hierarchical systems consisting of pure non-degenerate binaries and systems hosting evolved non-degenerate companions that have not undergone supernova events. This distinction reflects a fundamental physical difference between the two samples. Within the secondary-priority candidates, systems with higher \texttt{fit\_companion\_mass} or comparatively higher $V_\mathrm{pec}$ are therefore comparatively more promising sources for follow-up observations. On the other hand, UV excess could also arise from the chromospheric activity of the RGB. Consequently, such systems may be removed by our SED diagnostic criterion. This is consistent with our goal of prioritizing sample purity over completeness.

\begin{deluxetable*}{ccccccc}
\setlength{\tabcolsep}{2.5pt}
\tablewidth{0pt}
\tablecaption{Priority and secondary-priority candidates \label{tab:candidates}}
\tablehead{
\colhead{\emph{Gaia} DR3} & \colhead{\( \chi^2 \)/$N$} & \colhead{$R_\mathrm{NUV}$} & \colhead{$F_\mathrm{NUV,tot\_mod}$/$F_\mathrm{NUV,obs}$}  & \colhead{$V_\mathrm{pec}$} & \colhead{$|Z|_\mathrm{max}$} & \colhead{Fit\_comp\_mass} \\
\colhead{ID} & \colhead{} & \colhead{} & \colhead{} & \colhead{(km s$^{-1}$)} & \colhead{(kpc)} & \colhead{($M_\odot$)}}
\startdata
413801205418517760 & 0.90 & 0.93$\pm$0.14 & &
351 & 18.8 & 9.85$^{+6.60}_{-3.95}$ \\
4569884068106889088 & 1.02 & 2.99$\pm$0.24 & 2550$^{+220}_{-190}$ & 180 & 3.1 & 8.10$^{+5.90}_{-3.41}$ \\
2799283010053799936 & 1.20 & 1.59$\pm$0.19 & 6530$^{+880}_{-690}$ & 248 & 6.6 & 6.80$^{+6.58}_{-3.34}$ \\
4348345531815266688 & 1.27 & 1.32$\pm$0.17 & 26400$^{+4000}_{-3100}$ & 116 & 1.1 & 6.61$^{+5.15}_{-2.89}$ \\
6215201892206723712 & 1.19 & 2.59$\pm$0.52 & 3030$^{+770}_{-510}$ & 163 & 2.0 & 5.72$^{+5.12}_{-2.70}$ \\
4370292986492417280 & 1.08 & 1.05$\pm$0.52 & &
134 & 0.8 & 5.60$^{+5.14}_{-2.68}$ \\
5189794217106440192 & 1.63 & 1.31$\pm$0.09 & 2460$^{+190}_{-160}$ & 126 & 1.4 & 5.12$^{+5.67}_{-2.69}$ \\
4696396379961787520 & 1.38 & 1.54$\pm$0.30 & 930$^{+230}_{-150}$ & 228 & 6.5 & 4.75$^{+7.00}_{-2.83}$ \\
2431937996280114304 & 0.47 & 1.73$\pm$0.09 & 684$^{+36}_{-33}$ & 127 & 2.9 & 4.57$^{+4.72}_{-2.32}$ \\
4775580703429027328 & 0.43 & 0.66$\pm$0.08 & &
163 & 3.9 & 4.53$^{+3.72}_{-2.04}$ \\
6082713592919774336 & 0.71 & 0.77$\pm$0.20 & &
122 & 1.2 & 4.47$^{+4.48}_{-2.24}$ \\
6857124646247102720 & 1.70 & 2.72$\pm$0.45 & 2550$^{+500}_{-360}$ & 130 & 2.6 & 4.25$^{+3.08}_{-1.79}$ \\
6903875257889043200 & 0.70 & 1.27$\pm$0.21 & 590$^{+120}_{-90}$ & 395 & 19.0 & 4.17$^{+4.99}_{-2.27}$ \\
6461298879698884736 & 0.74 & 1.76$\pm$0.15 & 4060$^{+390}_{-330}$ & 123 & 1.3 & 3.96$^{+5.46}_{-2.29}$ \\
5690035586423107072 & 1.38 & 2.17$\pm$0.15 & 1240$^{+90}_{-80}$ & 109 & 0.9 & 3.51$^{+3.60}_{-1.78}$ \\
984731689102817408 & 0.75 & 1.70$\pm$0.19 & 1280$^{+160}_{-130}$ & 225 & 2.9 & 3.49$^{+5.17}_{-2.08}$ \\
830954817485226240 & 0.49 & 0.60$\pm$0.03 & &
132 & 2.2 & 3.43$^{+2.66}_{-1.50}$ \\
2689755154957920896 & 0.88 & 1.11$\pm$0.06 & &
122 & 3.1 & 3.38$^{+2.77}_{-1.52}$ \\
4624140465810214784 & 0.65 & 1.13$\pm$0.14 & &
120 & 3.1 & 3.29$^{+4.63}_{-1.92}$ \\
... & ... & ... & ... & ... & ... & ... \\
\enddata
    \begin{tablenotes}
        \item[1] Notes: column(1): \emph{Gaia} DR3 solution ID; column(2): \( \chi^2 \) per data point; column(3): the ratio of observed NUV flux to model-predicted NUV flux (the Y-axis in Figure \ref{fig:SED_diagnostics} (a)); column(4): the ratio of total model-predicted NUV flux (assuming MS companion) to observed NUV flux (the Y-axis in Figure \ref{fig:SED_diagnostics} (b)); column(5): Peculiar velocity calculated in Section \ref{sec4}; column(6): The maximum vertical height above/below the Galactic plane calculated in Section \ref{sec4}; column(7): The fitted companion mass provided by \citet{2025arXiv251005982M}. Only 19 highest-priority BH candidates are listed here and this table is available in its entirety in machine-readable form in the online article.       
    \end{tablenotes}
\end{deluxetable*}

\begin{acknowledgments}
We thank Jifeng Liu, Yang Huang, and Zhi-Xiang Zhang for helpful discussion, and the anonymous referee for constructive suggestions that improved the paper. This work was supported by the National Key R\&D Program of China under grants 2023YFA1607901 and 2021YFA1600401, the National Natural Science Foundation of China under grants 12433007 and 12221003. We acknowledge the science research grants from the China Manned Space Project with No. CMS-CSST-2025-A13. This work has made use of data from the European Space Agency (ESA) mission Gaia (\url{https://www.cosmos.esa.int/gaia}), processed by the Gaia Data Processing and Analysis Consortium (DPAC, \url{https://www.cosmos.esa.int/web/gaia/dpac/consortium}). Funding for the DPAC has been provided by national institutions, in particular the institutions participating in the Gaia Multilateral Agreement. This publication makes use of data products from the Two Micron All Sky Survey, which is a joint project of the University of Massachusetts and the Infrared Processing and Analysis Center/California Institute of Technology, funded by the National Aeronautics and Space Administration and the National Science Foundation. Database access and other data services are provided by the Associação Laboratório Interinstitucional de e-Astronomia (LIneA) with the financial support from INCT do e-Universo (Processo No. 465376/2014-2). This work makes use of GALEX and 2MASS. This publication makes use of VOSA, developed under the Spanish Virtual Observatory (\url{https://svo.cab.inta-csic.es}) project funded by MCIN/AEI/10.13039/501100011033/ through grant PID2020-112949GB-I00.
VOSA has been partially updated by using funding from the European Union's Horizon 2020 Research and Innovation Programme, under Grant Agreement No. 776403 (EXOPLANETS-A). This research has made use of the VizieR catalogue access tool, CDS, Strasbourg, France (DOI : 10.26093/cds/vizier). The original description of the VizieR service was published in 2000, A\&AS 143, 23. 
\end{acknowledgments}

\software{
          Astropy \citep{2013A&A...558A..33A, 2018AJ....156..123A, 2022ApJ...935..167A},
          astroquery
          \citep{2019AJ....157...98G},
          Numpy \citep{harris2020array}, 
          Pandas \citep{reback2020pandas},
          matplotlib \citep{Hunter:2007},
          emcee \citep{2013PASP..125..306F},
          Vizier \citep{2000A&AS..143...23O}, galpy \citep{2015ApJS..216...29B}, VOSA
          \citep{2008A&A...492..277B},
          extinction
          \citep{2016zndo....804967B},
          Isochrones
          \citep{2015ascl.soft03010M},
          pystellibs          (\url{https://github.com/mfouesneau/pystellibs}),
          spectool
          \citep{2025zndo..14947417Z}
          }

\bibliography{ref}{}

@ARTICLE{2025arXiv251005982M,
       author = {{M{\"u}ller-Horn}, Johanna and {Rix}, Hans-Walter and {El-Badry}, Kareem and {Pennell}, Ben and {Green}, Matthew and {Li}, Jiadong and {Seeburger}, Rhys},
        title = "{Dormant BH candidates from Gaia DR3 summary diagnostics}",
      journal = {arXiv e-prints},
         year = 2025,
        month = oct,
          eid = {arXiv:2510.05982},
        pages = {arXiv:2510.05982},
          doi = {10.48550/arXiv.2510.05982},
archivePrefix = {arXiv},
       eprint = {2510.05982},
 primaryClass = {astro-ph.SR},
       adsurl = {https://ui.adsabs.harvard.edu/abs/2025arXiv251005982M}
}

@ARTICLE{2010A&A...510A..23S,
       author = {{Sartore}, N. and {Ripamonti}, E. and {Treves}, A. and {Turolla}, R.},
        title = "{Galactic neutron stars. I. Space and velocity distributions in the disk and in the halo}",
      journal = {\aap},
         year = 2010,
        month = feb,
       volume = {510},
          eid = {A23},
        pages = {A23},
          doi = {10.1051/0004-6361/200912222},
archivePrefix = {arXiv},
       eprint = {0908.3182},
 primaryClass = {astro-ph.GA},
       adsurl = {https://ui.adsabs.harvard.edu/abs/2010A&A...510A..23S}
}

@ARTICLE{1996ApJ...457..834T,
       author = {{Timmes}, F.~X. and {Woosley}, S.~E. and {Weaver}, Thomas A.},
        title = "{The Neutron Star and Black Hole Initial Mass Function}",
      journal = {\apj},
         year = 1996,
        month = feb,
       volume = {457},
        pages = {834},
          doi = {10.1086/176778},
archivePrefix = {arXiv},
       eprint = {astro-ph/9510136},
 primaryClass = {astro-ph},
       adsurl = {https://ui.adsabs.harvard.edu/abs/1996ApJ...457..834T}
}

@ARTICLE{2024A&A...684A.124F,
       author = {{Fortin}, F. and {Kalsi}, A. and {Garc{\'\i}a}, F. and {Simaz-Bunzel}, A. and {Chaty}, S.},
        title = "{A catalogue of low-mass X-ray binaries in the Galaxy: From the INTEGRAL to the Gaia era}",
      journal = {\aap},
         year = 2024,
        month = apr,
       volume = {684},
          eid = {A124},
        pages = {A124},
          doi = {10.1051/0004-6361/202347908},
archivePrefix = {arXiv},
       eprint = {2401.11931},
 primaryClass = {astro-ph.HE},
       adsurl = {https://ui.adsabs.harvard.edu/abs/2024A&A...684A.124F}
}

@ARTICLE{2016A&A...587A..61C,
       author = {{Corral-Santana}, J.~M. and {Casares}, J. and {Mu{\~n}oz-Darias}, T. and {Bauer}, F.~E. and {Mart{\'\i}nez-Pais}, I.~G. and {Russell}, D.~M.},
        title = "{BlackCAT: A catalogue of stellar-mass black holes in X-ray transients}",
      journal = {\aap},
         year = 2016,
        month = mar,
       volume = {587},
          eid = {A61},
        pages = {A61},
          doi = {10.1051/0004-6361/201527130},
archivePrefix = {arXiv},
       eprint = {1510.08869},
 primaryClass = {astro-ph.HE},
       adsurl = {https://ui.adsabs.harvard.edu/abs/2016A&A...587A..61C}
}

@ARTICLE{2005AJ....129.1993M,
       author = {{Manchester}, R.~N. and {Hobbs}, G.~B. and {Teoh}, A. and {Hobbs}, M.},
        title = "{The Australia Telescope National Facility Pulsar Catalogue}",
      journal = {\aj},
         year = 2005,
        month = apr,
       volume = {129},
       number = {4},
        pages = {1993-2006},
          doi = {10.1086/428488},
archivePrefix = {arXiv},
       eprint = {astro-ph/0412641},
 primaryClass = {astro-ph},
       adsurl = {https://ui.adsabs.harvard.edu/abs/2005AJ....129.1993M}
}

@ARTICLE{2016MNRAS.458.3012W,
       author = {{Wyrzykowski}, {\L}. and {Kostrzewa-Rutkowska}, Z. and {Skowron}, J. and {Rybicki}, K.~A. and {Mr{\'o}z}, P. and {Koz{\l}owski}, S. and {Udalski}, A. and {Szyma{\'n}ski}, M.~K. and {Pietrzy{\'n}ski}, G. and {Soszy{\'n}ski}, I. and {Ulaczyk}, K. and {Pietrukowicz}, P. and {Poleski}, R. and {Pawlak}, M. and {I{\l}kiewicz}, K. and {Rattenbury}, N.~J.},
        title = "{Black hole, neutron star and white dwarf candidates from microlensing with OGLE-III}",
      journal = {\mnras},
         year = 2016,
        month = may,
       volume = {458},
       number = {3},
        pages = {3012-3026},
          doi = {10.1093/mnras/stw426},
archivePrefix = {arXiv},
       eprint = {1509.04899},
 primaryClass = {astro-ph.SR},
       adsurl = {https://ui.adsabs.harvard.edu/abs/2016MNRAS.458.3012W}
}

@ARTICLE{2025ApJ...983..104S,
       author = {{Sahu}, Kailash C. and {Anderson}, Jay and {Casertano}, Stefano and {Bond}, Howard E. and {Dominik}, Martin and {Calamida}, Annalisa and {Bellini}, Andrea and {Brown}, Thomas M. and {Ferguson}, Henry C. and {Rejkuba}, Marina},
        title = "{OGLE-2011-BLG-0462: An Isolated Stellar-mass Black Hole Confirmed Using New HST Astrometry and Updated Photometry}",
      journal = {\apj},
         year = 2025,
        month = apr,
       volume = {983},
       number = {2},
          eid = {104},
        pages = {104},
          doi = {10.3847/1538-4357/adbe6e},
archivePrefix = {arXiv},
       eprint = {2503.07820},
 primaryClass = {astro-ph.SR},
       adsurl = {https://ui.adsabs.harvard.edu/abs/2025ApJ...983..104S}
}

@ARTICLE{2016PhRvL.116f1102A,
       author = {{Abbott}, B.~P. and {Abbott}, R. and {Abbott}, T.~D. and {Abernathy}, M.~R. and {Acernese}, F. and {Ackley}, K. and {Adams}, C. and {Adams}, T. and {Addesso}, P. and {Adhikari}, R.~X. and {Adya}, V.~B. and {Affeldt}, C. and {Agathos}, M. and {Agatsuma}, K. and {Aggarwal}, N. and {Aguiar}, O.~D. and {Aiello}, L. and {Ain}, A. and {Ajith}, P. and {Allen}, B. and {Allocca}, A. and {Altin}, P.~A. and {Anderson}, S.~B. and {Anderson}, W.~G. and {Arai}, K. and {Arain}, M.~A. and {Araya}, M.~C. and {Arceneaux}, C.~C. and {Areeda}, J.~S. and {Arnaud}, N. and {Arun}, K.~G. and {Ascenzi}, S. and {Ashton}, G. and {Ast}, M. and {Aston}, S.~M. and {Astone}, P. and {Aufmuth}, P. and {Aulbert}, C. and {Babak}, S. and {Bacon}, P. and {Bader}, M.~K.~M. and {Baker}, P.~T. and {Baldaccini}, F. and {Ballardin}, G. and {Ballmer}, S.~W. and {Barayoga}, J.~C. and {Barclay}, S.~E. and {Barish}, B.~C. and {Barker}, D. and {Barone}, F. and {Barr}, B. and {Barsotti}, L. and {Barsuglia}, M. and {Barta}, D. and {Bartlett}, J. and {Barton}, M.~A. and {Bartos}, I. and {Bassiri}, R. and {Basti}, A. and {Batch}, J.~C. and {Baune}, C. and {Bavigadda}, V. and {Bazzan}, M. and {Behnke}, B. and {Bejger}, M. and {Belczynski}, C. and {Bell}, A.~S. and {Bell}, C.~J. and {Berger}, B.~K. and {Bergman}, J. and {Bergmann}, G. and {Berry}, C.~P.~L. and {Bersanetti}, D. and {Bertolini}, A. and {Betzwieser}, J. and {Bhagwat}, S. and {Bhandare}, R. and {Bilenko}, I.~A. and {Billingsley}, G. and {Birch}, J. and {Birney}, I.~A. and {Birnholtz}, O. and {Biscans}, S. and {Bisht}, A. and {Bitossi}, M. and {Biwer}, C. and {Bizouard}, M.~A. and {Blackburn}, J.~K. and {Blair}, C.~D. and {Blair}, D.~G. and {Blair}, R.~M. and {Bloemen}, S. and {Bock}, O. and {Bodiya}, T.~P. and {Boer}, M. and {Bogaert}, G. and {Bogan}, C. and {Bohe}, A. and {Bojtos}, P. and {Bond}, C. and {Bondu}, F. and {Bonnand}, R. and {Boom}, B.~A. and {Bork}, R. and {Boschi}, V. and {Bose}, S. and {Bouffanais}, Y. and {Bozzi}, A. and {Bradaschia}, C. and {Brady}, P.~R. and {Braginsky}, V.~B. and {Branchesi}, M. and {Brau}, J.~E. and {Briant}, T. and {Brillet}, A. and {Brinkmann}, M. and {Brisson}, V. and {Brockill}, P. and {Brooks}, A.~F. and {Brown}, D.~A. and {Brown}, D.~D. and {Brown}, N.~M. and {Buchanan}, C.~C. and {Buikema}, A. and {Bulik}, T. and {Bulten}, H.~J. and {Buonanno}, A. and {Buskulic}, D. and {Buy}, C. and {Byer}, R.~L. and {Cabero}, M. and {Cadonati}, L. and {Cagnoli}, G. and {Cahillane}, C. and {Bustillo}, J. Calder{\'o}n and {Callister}, T. and {Calloni}, E. and {Camp}, J.~B. and {Cannon}, K.~C. and {Cao}, J. and {Capano}, C.~D. and {Capocasa}, E. and {Carbognani}, F. and {Caride}, S. and {Diaz}, J. Casanueva and {Casentini}, C. and {Caudill}, S. and {Cavagli{\`a}}, M. and {Cavalier}, F. and {Cavalieri}, R. and {Cella}, G. and {Cepeda}, C.~B. and {Baiardi}, L. Cerboni and {Cerretani}, G. and {Cesarini}, E. and {Chakraborty}, R. and {Chalermsongsak}, T. and {Chamberlin}, S.~J. and {Chan}, M. and {Chao}, S. and {Charlton}, P. and {Chassande-Mottin}, E. and {Chen}, H.~Y. and {Chen}, Y. and {Cheng}, C. and {Chincarini}, A. and {Chiummo}, A. and {Cho}, H.~S. and {Cho}, M. and {Chow}, J.~H. and {Christensen}, N. and {Chu}, Q. and {Chua}, S. and {Chung}, S. and {Ciani}, G. and {Clara}, F. and {Clark}, J.~A. and {Cleva}, F. and {Coccia}, E. and {Cohadon}, P.-F. and {Colla}, A. and {Collette}, C.~G. and {Cominsky}, L. and {Constancio}, M. and {Conte}, A. and {Conti}, L. and {Cook}, D. and {Corbitt}, T.~R. and {Cornish}, N. and {Corsi}, A. and {Cortese}, S. and {Costa}, C.~A. and {Coughlin}, M.~W. and {Coughlin}, S.~B. and {Coulon}, J.-P. and {Countryman}, S.~T. and {Couvares}, P. and {Cowan}, E.~E. and {Coward}, D.~M. and {Cowart}, M.~J.},
        title = "{Observation of Gravitational Waves from a Binary Black Hole Merger}",
      journal = {\prl},
         year = 2016,
        month = feb,
       volume = {116},
       number = {6},
          eid = {061102},
        pages = {061102},
          doi = {10.1103/PhysRevLett.116.061102},
archivePrefix = {arXiv},
       eprint = {1602.03837},
 primaryClass = {gr-qc},
       adsurl = {https://ui.adsabs.harvard.edu/abs/2016PhRvL.116f1102A}
}

@ARTICLE{2025arXiv250818079T,
       author = {{The LIGO Scientific Collaboration} and {the Virgo Collaboration} and {the KAGRA Collaboration} and {Abac}, A.~G. and {Abouelfettouh}, I. and {Acernese}, F. and {Ackley}, K. and {Adamcewicz}, C. and {Adhicary}, S. and {Adhikari}, D. and {Adhikari}, N. and {Adhikari}, R.~X. and {Adkins}, V.~K. and {Afroz}, S. and {Agapito}, A. and {Agarwal}, D. and {Agathos}, M. and {Aggarwal}, N. and {Aggarwal}, S. and {Aguiar}, O.~D. and {Ahrend}, I.-L. and {Aiello}, L. and {Ain}, A. and {Ajith}, P. and {Akutsu}, T. and {Albanesi}, S. and {Ali}, W. and {Al-Kershi}, S. and {All{\'e}n{\'e}}, C. and {Allocca}, A. and {Al-Shammari}, S. and {Altin}, P.~A. and {Alvarez-Lopez}, S. and {Amar}, W. and {Amarasinghe}, O. and {Amato}, A. and {Amicucci}, F. and {Amra}, C. and {Ananyeva}, A. and {Anderson}, S.~B. and {Anderson}, W.~G. and {Andia}, M. and {Ando}, M. and {Andr{\'e}s-Carcasona}, M. and {Andri{\'c}}, T. and {Anglin}, J. and {Ansoldi}, S. and {Antelis}, J.~M. and {Antier}, S. and {Aoumi}, M. and {Appavuravther}, E.~Z. and {Appert}, S. and {Apple}, S.~K. and {Arai}, K. and {Araya}, A. and {Araya}, M.~C. and {Arca Sedda}, M. and {Areeda}, J.~S. and {Aritomi}, N. and {Armato}, F. and {Armstrong}, S. and {Arnaud}, N. and {Arogeti}, M. and {Aronson}, S.~M. and {Ashton}, G. and {Aso}, Y. and {Asprea}, L. and {Assiduo}, M. and {Assis de Souza Melo}, S. and {Aston}, S.~M. and {Astone}, P. and {Attadio}, F. and {Aubin}, F. and {AultONeal}, K. and {Avallone}, G. and {Avila}, E.~A. and {Babak}, S. and {Badger}, C. and {Bae}, S. and {Bagnasco}, S. and {Baiotti}, L. and {Bajpai}, R. and {Baka}, T. and {Baker}, A.~M. and {Baker}, K.~A. and {Baker}, T. and {Baldi}, G. and {Baldicchi}, N. and {Ball}, M. and {Ballardin}, G. and {Ballmer}, S.~W. and {Banagiri}, S. and {Banerjee}, B. and {Bankar}, D. and {Baptiste}, T.~M. and {Baral}, P. and {Baratti}, M. and {Barayoga}, J.~C. and {Barish}, B.~C. and {Barker}, D. and {Barman}, N. and {Barneo}, P. and {Barone}, F. and {Barr}, B. and {Barsotti}, L. and {Barsuglia}, M. and {Barta}, D. and {Bartoletti}, A.~M. and {Barton}, M.~A. and {Bartos}, I. and {Basalaev}, A. and {Bassiri}, R. and {Basti}, A. and {Bawaj}, M. and {Baxi}, P. and {Bayley}, J.~C. and {Baylor}, A.~C. and {Baynard}, II, P.~A. and {Bazzan}, M. and {Bedakihale}, V.~M. and {Beirnaert}, F. and {Bejger}, M. and {Belardinelli}, D. and {Bell}, A.~S. and {Bellie}, D.~S. and {Bellizzi}, L. and {Benoit}, W. and {Bentara}, I. and {Bentley}, J.~D. and {Ben Yaala}, M. and {Bera}, S. and {Bergamin}, F. and {Berger}, B.~K. and {Bernuzzi}, S. and {Beroiz}, M. and {Berry}, C.~P.~L. and {Bersanetti}, D. and {Bertheas}, T. and {Bertolini}, A. and {Betzwieser}, J. and {Beveridge}, D. and {Bevilacqua}, G. and {Bevins}, N. and {Bhandare}, R. and {Bhatt}, R. and {Bhattacharjee}, D. and {Bhattacharyya}, S. and {Bhaumik}, S. and {Biancalana}, V. and {Bianchi}, A. and {Bilenko}, I.~A. and {Billingsley}, G. and {Binetti}, A. and {Bini}, S. and {Binu}, C. and {Biot}, S. and {Birnholtz}, O. and {Biscoveanu}, S. and {Bisht}, A. and {Bitossi}, M. and {Bizouard}, M.-A. and {Blaber}, S. and {Blackburn}, J.~K. and {Blagg}, L.~A. and {Blair}, C.~D. and {Blair}, D.~G. and {Bode}, N. and {Boettner}, N. and {Boileau}, G. and {Boldrini}, M. and {Bolingbroke}, G.~N. and {Bolliand}, A. and {Bonavena}, L.~D. and {Bondarescu}, R. and {Bondu}, F. and {Bonilla}, E. and {Bonilla}, M.~S. and {Bonino}, A. and {Bonnand}, R. and {Borchers}, A. and {Borhanian}, S. and {Boschi}, V. and {Bose}, S. and {Bossilkov}, V. and {Bothra}, Y. and {Boudon}, A. and {Bourg}, L. and {Boyle}, M. and {Bozzi}, A. and {Bradaschia}, C. and {Brady}, P.~R. and {Branch}, A. and {Branchesi}, M. and {Braun}, I. and {Briant}, T. and {Brillet}, A. and {Brinkmann}, M. and {Brockill}, P. and {Brockmueller}, E. and {Brooks}, A.~F.},
        title = "{Open Data from LIGO, Virgo, and KAGRA through the First Part of the Fourth Observing Run}",
      journal = {arXiv e-prints},
         year = 2025,
        month = aug,
          eid = {arXiv:2508.18079},
        pages = {arXiv:2508.18079},
          doi = {10.48550/arXiv.2508.18079},
archivePrefix = {arXiv},
       eprint = {2508.18079},
 primaryClass = {gr-qc},
       adsurl = {https://ui.adsabs.harvard.edu/abs/2025arXiv250818079T}
}

@ARTICLE{2022NatAs...6.1203Y,
       author = {{Yi}, Tuan and {Gu}, Wei-Min and {Zhang}, Zhi-Xiang and {Zheng}, Ling-Lin and {Sun}, Mouyuan and {Wang}, Junfeng and {Bai}, Zhongrui and {Wang}, Pei and {Wu}, Jianfeng and {Bai}, Yu and {Wang}, Song and {Zhang}, Haotong and {Dong}, Yize and {Shao}, Yong and {Li}, Xiang-Dong and {Zhang}, Jia and {Huang}, Yang and {Yang}, Fan and {Yu}, Qingzheng and {Mu}, Hui-Jun and {Fu}, Jin-Bo and {Qi}, Senyu and {Guo}, Jing and {Fang}, Xuan and {Zheng}, Chuanjie and {Li}, Chun-Qian and {Shi}, Jian-Rong and {Chen}, Huanyang and {Liu}, Jifeng},
        title = "{A dynamically discovered and characterized non-accreting neutron star-M dwarf binary candidate}",
      journal = {Nature Astronomy},
         year = 2022,
        month = sep,
       volume = {6},
        pages = {1203-1212},
          doi = {10.1038/s41550-022-01766-0},
archivePrefix = {arXiv},
       eprint = {2209.12141},
 primaryClass = {astro-ph.SR},
       adsurl = {https://ui.adsabs.harvard.edu/abs/2022NatAs...6.1203Y}
}

@ARTICLE{2022A&A...664A.159M,
       author = {{Mahy}, L. and {Sana}, H. and {Shenar}, T. and {Sen}, K. and {Langer}, N. and {Marchant}, P. and {Abdul-Masih}, M. and {Banyard}, G. and {Bodensteiner}, J. and {Bowman}, D.~M. and {Dsilva}, K. and {Fabry}, M. and {Hawcroft}, C. and {Janssens}, S. and {Van Reeth}, T. and {Eldridge}, C.},
        title = "{Identifying quiescent compact objects in massive Galactic single-lined spectroscopic binaries}",
      journal = {\aap},
         year = 2022,
        month = aug,
       volume = {664},
          eid = {A159},
        pages = {A159},
          doi = {10.1051/0004-6361/202243147},
archivePrefix = {arXiv},
       eprint = {2207.07752},
 primaryClass = {astro-ph.SR},
       adsurl = {https://ui.adsabs.harvard.edu/abs/2022A&A...664A.159M}
}

@ARTICLE{2023A&A...674A...1G,
       author = {{Gaia Collaboration} and {Vallenari}, A. and {Brown}, A.~G.~A. and {Prusti}, T. and {de Bruijne}, J.~H.~J. and {Arenou}, F. and {Babusiaux}, C. and {Biermann}, M. and {Creevey}, O.~L. and {Ducourant}, C. and {Evans}, D.~W. and {Eyer}, L. and {Guerra}, R. and {Hutton}, A. and {Jordi}, C. and {Klioner}, S.~A. and {Lammers}, U.~L. and {Lindegren}, L. and {Luri}, X. and {Mignard}, F. and {Panem}, C. and {Pourbaix}, D. and {Randich}, S. and {Sartoretti}, P. and {Soubiran}, C. and {Tanga}, P. and {Walton}, N.~A. and {Bailer-Jones}, C.~A.~L. and {Bastian}, U. and {Drimmel}, R. and {Jansen}, F. and {Katz}, D. and {Lattanzi}, M.~G. and {van Leeuwen}, F. and {Bakker}, J. and {Cacciari}, C. and {Casta{\~n}eda}, J. and {De Angeli}, F. and {Fabricius}, C. and {Fouesneau}, M. and {Fr{\'e}mat}, Y. and {Galluccio}, L. and {Guerrier}, A. and {Heiter}, U. and {Masana}, E. and {Messineo}, R. and {Mowlavi}, N. and {Nicolas}, C. and {Nienartowicz}, K. and {Pailler}, F. and {Panuzzo}, P. and {Riclet}, F. and {Roux}, W. and {Seabroke}, G.~M. and {Sordo}, R. and {Th{\'e}venin}, F. and {Gracia-Abril}, G. and {Portell}, J. and {Teyssier}, D. and {Altmann}, M. and {Andrae}, R. and {Audard}, M. and {Bellas-Velidis}, I. and {Benson}, K. and {Berthier}, J. and {Blomme}, R. and {Burgess}, P.~W. and {Busonero}, D. and {Busso}, G. and {C{\'a}novas}, H. and {Carry}, B. and {Cellino}, A. and {Cheek}, N. and {Clementini}, G. and {Damerdji}, Y. and {Davidson}, M. and {de Teodoro}, P. and {Nu{\~n}ez Campos}, M. and {Delchambre}, L. and {Dell'Oro}, A. and {Esquej}, P. and {Fern{\'a}ndez-Hern{\'a}ndez}, J. and {Fraile}, E. and {Garabato}, D. and {Garc{\'\i}a-Lario}, P. and {Gosset}, E. and {Haigron}, R. and {Halbwachs}, J.-L. and {Hambly}, N.~C. and {Harrison}, D.~L. and {Hern{\'a}ndez}, J. and {Hestroffer}, D. and {Hodgkin}, S.~T. and {Holl}, B. and {Jan{\ss}en}, K. and {Jevardat de Fombelle}, G. and {Jordan}, S. and {Krone-Martins}, A. and {Lanzafame}, A.~C. and {L{\"o}ffler}, W. and {Marchal}, O. and {Marrese}, P.~M. and {Moitinho}, A. and {Muinonen}, K. and {Osborne}, P. and {Pancino}, E. and {Pauwels}, T. and {Recio-Blanco}, A. and {Reyl{\'e}}, C. and {Riello}, M. and {Rimoldini}, L. and {Roegiers}, T. and {Rybizki}, J. and {Sarro}, L.~M. and {Siopis}, C. and {Smith}, M. and {Sozzetti}, A. and {Utrilla}, E. and {van Leeuwen}, M. and {Abbas}, U. and {{\'A}brah{\'a}m}, P. and {Abreu Aramburu}, A. and {Aerts}, C. and {Aguado}, J.~J. and {Ajaj}, M. and {Aldea-Montero}, F. and {Altavilla}, G. and {{\'A}lvarez}, M.~A. and {Alves}, J. and {Anders}, F. and {Anderson}, R.~I. and {Anglada Varela}, E. and {Antoja}, T. and {Baines}, D. and {Baker}, S.~G. and {Balaguer-N{\'u}{\~n}ez}, L. and {Balbinot}, E. and {Balog}, Z. and {Barache}, C. and {Barbato}, D. and {Barros}, M. and {Barstow}, M.~A. and {Bartolom{\'e}}, S. and {Bassilana}, J.-L. and {Bauchet}, N. and {Becciani}, U. and {Bellazzini}, M. and {Berihuete}, A. and {Bernet}, M. and {Bertone}, S. and {Bianchi}, L. and {Binnenfeld}, A. and {Blanco-Cuaresma}, S. and {Blazere}, A. and {Boch}, T. and {Bombrun}, A. and {Bossini}, D. and {Bouquillon}, S. and {Bragaglia}, A. and {Bramante}, L. and {Breedt}, E. and {Bressan}, A. and {Brouillet}, N. and {Brugaletta}, E. and {Bucciarelli}, B. and {Burlacu}, A. and {Butkevich}, A.~G. and {Buzzi}, R. and {Caffau}, E. and {Cancelliere}, R. and {Cantat-Gaudin}, T. and {Carballo}, R. and {Carlucci}, T. and {Carnerero}, M.~I. and {Carrasco}, J.~M. and {Casamiquela}, L. and {Castellani}, M. and {Castro-Ginard}, A. and {Chaoul}, L. and {Charlot}, P. and {Chemin}, L. and {Chiaramida}, V. and {Chiavassa}, A. and {Chornay}, N. and {Comoretto}, G. and {Contursi}, G. and {Cooper}, W.~J. and {Cornez}, T. and {Cowell}, S. and {Crifo}, F. and {Cropper}, M. and {Crosta}, M. and {Crowley}, C. and {Dafonte}, C. and {Dapergolas}, A. and {David}, M. and {David}, P. and {de Laverny}, P. and {De Luise}, F. and {De March}, R.},
        title = "{Gaia Data Release 3. Summary of the content and survey properties}",
      journal = {\aap},
         year = 2023,
        month = jun,
       volume = {674},
          eid = {A1},
        pages = {A1},
          doi = {10.1051/0004-6361/202243940},
archivePrefix = {arXiv},
       eprint = {2208.00211},
 primaryClass = {astro-ph.GA},
       adsurl = {https://ui.adsabs.harvard.edu/abs/2023A&A...674A...1G}
}

@ARTICLE{2023MNRAS.518.1057E,
       author = {{El-Badry}, Kareem and {Rix}, Hans-Walter and {Quataert}, Eliot and {Howard}, Andrew W. and {Isaacson}, Howard and {Fuller}, Jim and {Hawkins}, Keith and {Breivik}, Katelyn and {Wong}, Kaze W.~K. and {Rodriguez}, Antonio C. and {Conroy}, Charlie and {Shahaf}, Sahar and {Mazeh}, Tsevi and {Arenou}, Fr{\'e}d{\'e}ric and {Burdge}, Kevin B. and {Bashi}, Dolev and {Faigler}, Simchon and {Weisz}, Daniel R. and {Seeburger}, Rhys and {Almada Monter}, Silvia and {Wojno}, Jennifer},
        title = "{A Sun-like star orbiting a black hole}",
      journal = {\mnras},
         year = 2023,
        month = jan,
       volume = {518},
       number = {1},
        pages = {1057-1085},
          doi = {10.1093/mnras/stac3140},
archivePrefix = {arXiv},
       eprint = {2209.06833},
 primaryClass = {astro-ph.SR},
       adsurl = {https://ui.adsabs.harvard.edu/abs/2023MNRAS.518.1057E}
}

@ARTICLE{2023MNRAS.521.4323E,
       author = {{El-Badry}, Kareem and {Rix}, Hans-Walter and {Cendes}, Yvette and {Rodriguez}, Antonio C. and {Conroy}, Charlie and {Quataert}, Eliot and {Hawkins}, Keith and {Zari}, Eleonora and {Hobson}, Melissa and {Breivik}, Katelyn and {Rau}, Arne and {Berger}, Edo and {Shahaf}, Sahar and {Seeburger}, Rhys and {Burdge}, Kevin B. and {Latham}, David W. and {Buchhave}, Lars A. and {Bieryla}, Allyson and {Bashi}, Dolev and {Mazeh}, Tsevi and {Faigler}, Simchon},
        title = "{A red giant orbiting a black hole}",
      journal = {\mnras},
         year = 2023,
        month = may,
       volume = {521},
       number = {3},
        pages = {4323-4348},
          doi = {10.1093/mnras/stad799},
archivePrefix = {arXiv},
       eprint = {2302.07880},
 primaryClass = {astro-ph.SR},
       adsurl = {https://ui.adsabs.harvard.edu/abs/2023MNRAS.521.4323E}
}

@ARTICLE{2024NatAs...8.1583W,
       author = {{Wang}, Song and {Zhao}, Xinlin and {Feng}, Fabo and {Ge}, Hongwei and {Shao}, Yong and {Cui}, Yingzhen and {Gao}, Shijie and {Zhang}, Lifu and {Wang}, Pei and {Li}, Xue and {Bai}, Zhongrui and {Yuan}, Hailong and {Huang}, Yang and {Yuan}, Haibo and {Zhang}, Zhixiang and {Yi}, Tuan and {Xiang}, Maosheng and {Li}, Zhenwei and {Li}, Tanda and {Zhang}, Junbo and {Zhang}, Meng and {Han}, Henggeng and {Fan}, Dongwei and {Li}, Xiangdong and {Chen}, Xuefei and {Liu}, Zhengwei and {Meng}, Xiangcun and {Liu}, Qingzhong and {Zhang}, Haotong and {Gu}, Wei-Min and {Liu}, Jifeng},
        title = "{A potential mass-gap black hole in a wide binary with a circular orbit}",
      journal = {Nature Astronomy},
         year = 2024,
        month = dec,
       volume = {8},
        pages = {1583-1591},
          doi = {10.1038/s41550-024-02359-9},
archivePrefix = {arXiv},
       eprint = {2409.06352},
 primaryClass = {astro-ph.SR},
       adsurl = {https://ui.adsabs.harvard.edu/abs/2024NatAs...8.1583W}
}

@ARTICLE{2019Sci...366..637T,
       author = {{Thompson}, Todd A. and {Kochanek}, Christopher S. and {Stanek}, Krzysztof Z. and {Badenes}, Carles and {Post}, Richard S. and {Jayasinghe}, Tharindu and {Latham}, David W. and {Bieryla}, Allyson and {Esquerdo}, Gilbert A. and {Berlind}, Perry and {Calkins}, Michael L. and {Tayar}, Jamie and {Lindegren}, Lennart and {Johnson}, Jennifer A. and {Holoien}, Thomas W.-S. and {Auchettl}, Katie and {Covey}, Kevin},
        title = "{A noninteracting low-mass black hole-giant star binary system}",
      journal = {Science},
         year = 2019,
        month = nov,
       volume = {366},
       number = {6465},
        pages = {637-640},
          doi = {10.1126/science.aau4005},
archivePrefix = {arXiv},
       eprint = {1806.02751},
 primaryClass = {astro-ph.HE},
       adsurl = {https://ui.adsabs.harvard.edu/abs/2019Sci...366..637T}
}

@ARTICLE{2021MNRAS.504.2577J,
       author = {{Jayasinghe}, T. and {Stanek}, K.~Z. and {Thompson}, Todd A. and {Kochanek}, C.~S. and {Rowan}, D.~M. and {Vallely}, P.~J. and {Strassmeier}, K.~G. and {Weber}, M. and {Hinkle}, J.~T. and {Hambsch}, F.-J. and {Martin}, D.~V. and {Prieto}, J.~L. and {Pessi}, T. and {Huber}, D. and {Auchettl}, K. and {Lopez}, L.~A. and {Ilyin}, I. and {Badenes}, C. and {Howard}, A.~W. and {Isaacson}, H. and {Murphy}, S.~J.},
        title = "{A unicorn in monoceros: the 3 M$_{☉}$ dark companion to the bright, nearby red giant V723 Mon is a non-interacting, mass-gap black hole candidate}",
      journal = {\mnras},
         year = 2021,
        month = jun,
       volume = {504},
       number = {2},
        pages = {2577-2602},
          doi = {10.1093/mnras/stab907},
archivePrefix = {arXiv},
       eprint = {2101.02212},
 primaryClass = {astro-ph.SR},
       adsurl = {https://ui.adsabs.harvard.edu/abs/2021MNRAS.504.2577J}
}

@ARTICLE{2024ApJ...976..131B,
       author = {{Bianchi}, Luciana and {Hutchings}, John and {Bohlin}, Ralph and {Thilker}, David and {Berti}, Emanuele},
        title = "{Revealing the Elusive Companion of the Red Giant Binary 2MASS J05215658+4359220 from UV HST and Astrosat/UVIT Data}",
      journal = {\apj},
         year = 2024,
        month = nov,
       volume = {976},
       number = {1},
          eid = {131},
        pages = {131},
          doi = {10.3847/1538-4357/ad712f},
archivePrefix = {arXiv},
       eprint = {2409.06906},
 primaryClass = {astro-ph.SR},
       adsurl = {https://ui.adsabs.harvard.edu/abs/2024ApJ...976..131B}
}

@ARTICLE{2025arXiv250910608K,
       author = {{Kochanek}, C.~S. and {Stanek}, K.~Z. and {Thompson}, T.~A. and {Jayasinghe}, T.},
        title = "{The HST Ultraviolet Spectrum of V723 Mon: Additional Evidence of a Stellar Companion}",
      journal = {arXiv e-prints},
         year = 2025,
        month = sep,
          eid = {arXiv:2509.10608},
        pages = {arXiv:2509.10608},
          doi = {10.48550/arXiv.2509.10608},
archivePrefix = {arXiv},
       eprint = {2509.10608},
 primaryClass = {astro-ph.SR},
       adsurl = {https://ui.adsabs.harvard.edu/abs/2025arXiv250910608K}
}

@ARTICLE{2005ApJ...619L...1M,
       author = {{Martin}, D. Christopher and {Fanson}, James and {Schiminovich}, David and {Morrissey}, Patrick and {Friedman}, Peter G. and {Barlow}, Tom A. and {Conrow}, Tim and {Grange}, Robert and {Jelinsky}, Patrick N. and {Milliard}, Bruno and {Siegmund}, Oswald H.~W. and {Bianchi}, Luciana and {Byun}, Yong-Ik and {Donas}, Jose and {Forster}, Karl and {Heckman}, Timothy M. and {Lee}, Young-Wook and {Madore}, Barry F. and {Malina}, Roger F. and {Neff}, Susan G. and {Rich}, R. Michael and {Small}, Todd and {Surber}, Frank and {Szalay}, Alex S. and {Welsh}, Barry and {Wyder}, Ted K.},
        title = "{The Galaxy Evolution Explorer: A Space Ultraviolet Survey Mission}",
      journal = {\apjl},
         year = 2005,
        month = jan,
       volume = {619},
       number = {1},
        pages = {L1-L6},
          doi = {10.1086/426387},
archivePrefix = {arXiv},
       eprint = {astro-ph/0411302},
 primaryClass = {astro-ph},
       adsurl = {https://ui.adsabs.harvard.edu/abs/2005ApJ...619L...1M}
}

@ARTICLE{2005ApJ...619L...7M,
       author = {{Morrissey}, Patrick and {Schiminovich}, David and {Barlow}, Tom A. and {Martin}, D. Christopher and {Blakkolb}, Brian and {Conrow}, Tim and {Cooke}, Brian and {Erickson}, Kerry and {Fanson}, James and {Friedman}, Peter G. and {Grange}, Robert and {Jelinsky}, Patrick N. and {Lee}, Siu-Chun and {Liu}, Dankai and {Mazer}, Alan and {McLean}, Ryan and {Milliard}, Bruno and {Randall}, David and {Schmitigal}, Wes and {Sen}, Amit and {Siegmund}, Oswald H.~W. and {Surber}, Frank and {Vaughan}, Arthur and {Viton}, Maurice and {Welsh}, Barry Y. and {Bianchi}, Luciana and {Byun}, Yong-Ik and {Donas}, Jose and {Forster}, Karl and {Heckman}, Timothy M. and {Lee}, Young-Wook and {Madore}, Barry F. and {Malina}, Roger F. and {Neff}, Susan G. and {Rich}, R. Michael and {Small}, Todd and {Szalay}, Alex S. and {Wyder}, Ted K.},
        title = "{The On-Orbit Performance of the Galaxy Evolution Explorer}",
      journal = {\apjl},
         year = 2005,
        month = jan,
       volume = {619},
       number = {1},
        pages = {L7-L10},
          doi = {10.1086/424734},
archivePrefix = {arXiv},
       eprint = {astro-ph/0411310},
 primaryClass = {astro-ph},
       adsurl = {https://ui.adsabs.harvard.edu/abs/2005ApJ...619L...7M}
}

@INPROCEEDINGS{2015AAS...22533616H,
       author = {{Henden}, Arne A. and {Levine}, Stephen and {Terrell}, Dirk and {Welch}, Douglas L.},
        title = "{APASS - The Latest Data Release}",
    booktitle = {American Astronomical Society Meeting Abstracts \#225},
         year = 2015,
       series = {American Astronomical Society Meeting Abstracts},
       volume = {225},
        month = jan,
          eid = {336.16},
        pages = {336.16},
       adsurl = {https://ui.adsabs.harvard.edu/abs/2015AAS...22533616H}
}

@ARTICLE{2006AJ....131.1163S,
       author = {{Skrutskie}, M.~F. and {Cutri}, R.~M. and {Stiening}, R. and {Weinberg}, M.~D. and {Schneider}, S. and {Carpenter}, J.~M. and {Beichman}, C. and {Capps}, R. and {Chester}, T. and {Elias}, J. and {Huchra}, J. and {Liebert}, J. and {Lonsdale}, C. and {Monet}, D.~G. and {Price}, S. and {Seitzer}, P. and {Jarrett}, T. and {Kirkpatrick}, J.~D. and {Gizis}, J.~E. and {Howard}, E. and {Evans}, T. and {Fowler}, J. and {Fullmer}, L. and {Hurt}, R. and {Light}, R. and {Kopan}, E.~L. and {Marsh}, K.~A. and {McCallon}, H.~L. and {Tam}, R. and {Van Dyk}, S. and {Wheelock}, S.},
        title = "{The Two Micron All Sky Survey (2MASS)}",
      journal = {\aj},
         year = 2006,
        month = feb,
       volume = {131},
       number = {2},
        pages = {1163-1183},
          doi = {10.1086/498708},
       adsurl = {https://ui.adsabs.harvard.edu/abs/2006AJ....131.1163S}
}

@ARTICLE{2010AJ....140.1868W,
       author = {{Wright}, Edward L. and {Eisenhardt}, Peter R.~M. and {Mainzer}, Amy K. and {Ressler}, Michael E. and {Cutri}, Roc M. and {Jarrett}, Thomas and {Kirkpatrick}, J. Davy and {Padgett}, Deborah and {McMillan}, Robert S. and {Skrutskie}, Michael and {Stanford}, S.~A. and {Cohen}, Martin and {Walker}, Russell G. and {Mather}, John C. and {Leisawitz}, David and {Gautier}, III, Thomas N. and {McLean}, Ian and {Benford}, Dominic and {Lonsdale}, Carol J. and {Blain}, Andrew and {Mendez}, Bryan and {Irace}, William R. and {Duval}, Valerie and {Liu}, Fengchuan and {Royer}, Don and {Heinrichsen}, Ingolf and {Howard}, Joan and {Shannon}, Mark and {Kendall}, Martha and {Walsh}, Amy L. and {Larsen}, Mark and {Cardon}, Joel G. and {Schick}, Scott and {Schwalm}, Mark and {Abid}, Mohamed and {Fabinsky}, Beth and {Naes}, Larry and {Tsai}, Chao-Wei},
        title = "{The Wide-field Infrared Survey Explorer (WISE): Mission Description and Initial On-orbit Performance}",
      journal = {\aj},
         year = 2010,
        month = dec,
       volume = {140},
       number = {6},
        pages = {1868-1881},
          doi = {10.1088/0004-6256/140/6/1868},
archivePrefix = {arXiv},
       eprint = {1008.0031},
 primaryClass = {astro-ph.IM},
       adsurl = {https://ui.adsabs.harvard.edu/abs/2010AJ....140.1868W}
}

@ARTICLE{2025ApJ...984...58H,
       author = {{Huson}, Dylan and {Cowan}, Indiana and {Sizemore}, Logan and {Kounkel}, Marina and {Hutchinson}, Brian},
        title = "{Gaia Net: Toward Robust Spectroscopic Parameters of Stars of all Evolutionary Stages}",
      journal = {\apj},
         year = 2025,
        month = may,
       volume = {984},
       number = {1},
          eid = {58},
        pages = {58},
          doi = {10.3847/1538-4357/adc2fa},
archivePrefix = {arXiv},
       eprint = {2503.02958},
 primaryClass = {astro-ph.SR},
       adsurl = {https://ui.adsabs.harvard.edu/abs/2025ApJ...984...58H}
}

@ARTICLE{2013PASP..125..306F,
       author = {{Foreman-Mackey}, Daniel and {Hogg}, David W. and {Lang}, Dustin and {Goodman}, Jonathan},
        title = "{emcee: The MCMC Hammer}",
      journal = {\pasp},
         year = 2013,
        month = mar,
       volume = {125},
       number = {925},
        pages = {306},
          doi = {10.1086/670067},
archivePrefix = {arXiv},
       eprint = {1202.3665},
 primaryClass = {astro-ph.IM},
       adsurl = {https://ui.adsabs.harvard.edu/abs/2013PASP..125..306F}
}

@ARTICLE{2008A&A...492..277B,
       author = {{Bayo}, A. and {Rodrigo}, C. and {Barrado Y Navascu{\'e}s}, D. and {Solano}, E. and {Guti{\'e}rrez}, R. and {Morales-Calder{\'o}n}, M. and {Allard}, F.},
        title = "{VOSA: virtual observatory SED analyzer. An application to the Collinder 69 open cluster}",
      journal = {\aap},
         year = 2008,
        month = dec,
       volume = {492},
       number = {1},
        pages = {277-287},
          doi = {10.1051/0004-6361:200810395},
archivePrefix = {arXiv},
       eprint = {0808.0270},
 primaryClass = {astro-ph},
       adsurl = {https://ui.adsabs.harvard.edu/abs/2008A&A...492..277B}
}

@ARTICLE{2001ApJ...556..357A,
       author = {{Allard}, France and {Hauschildt}, Peter H. and {Alexander}, David R. and {Tamanai}, Akemi and {Schweitzer}, Andreas},
        title = "{The Limiting Effects of Dust in Brown Dwarf Model Atmospheres}",
      journal = {\apj},
         year = 2001,
        month = jul,
       volume = {556},
       number = {1},
        pages = {357-372},
          doi = {10.1086/321547},
archivePrefix = {arXiv},
       eprint = {astro-ph/0104256},
 primaryClass = {astro-ph},
       adsurl = {https://ui.adsabs.harvard.edu/abs/2001ApJ...556..357A}
}

@ARTICLE{2012RSPTA.370.2765A,
       author = {{Allard}, F. and {Homeier}, D. and {Freytag}, B.},
        title = "{Models of very-low-mass stars, brown dwarfs and exoplanets}",
      journal = {Philosophical Transactions of the Royal Society of London Series A},
         year = 2012,
        month = jun,
       volume = {370},
       number = {1968},
        pages = {2765-2777},
          doi = {10.1098/rsta.2011.0269},
archivePrefix = {arXiv},
       eprint = {1112.3591},
 primaryClass = {astro-ph.SR},
       adsurl = {https://ui.adsabs.harvard.edu/abs/2012RSPTA.370.2765A}
}
\bibliographystyle{aasjournalv7}

\end{document}